\documentclass[twocolumn]{aastex63}
\usepackage{amsmath}
\usepackage{mathtools}  
\usepackage{comment}
\usepackage{graphicx}
\usepackage{xcolor}
\usepackage{xparse,xcoffins}
\usepackage{appendix}

\defcitealias{2023MNRAS.525.5989Y}{Yang2023}
\shorttitle{Do cosmological reproduce $z \geq 12$ galaxy properties?}
\shortauthors{Yang et al.}

\ExplSyntaxOn
\NewCoffin\imagecoffin
\NewCoffin\labelcoffin

\keys_define:nn { miguel/label }
 {
  label   .tl_set:N = \l_miguel_label_tl,
  labelbox .bool_set:N = \l_miguel_label_box_bool,
  labelbox .default:n = true,
  fontsize .tl_set:N = \l_miguel_label_size_tl,
  fontsize .initial:n = \footnotesize,
  pos .choice:,
  pos/nw .code:n = \tl_set:Nn \l_miguel_label_pos_tl { left,up },
  pos/ne .code:n = \tl_set:Nn \l_miguel_label_pos_tl { right,up },
  pos/sw .code:n = \tl_set:Nn \l_miguel_label_pos_tl { left,down },
  pos/se .code:n = \tl_set:Nn \l_miguel_label_pos_tl { right,down },
  pos/n .code:n = \tl_set:Nn \l_miguel_label_pos_tl { hc,up },
  pos/w .code:n = \tl_set:Nn \l_miguel_label_pos_tl { left,vc },
  pos/s .code:n = \tl_set:Nn \l_miguel_label_pos_tl { hc,down },
  pos/e .code:n = \tl_set:Nn \l_miguel_label_pos_tl { right,vc },
  pos .initial:n = nw,
  unknown .code:n   = \clist_put_right:Nx \l_miguel_label_clist
                       { \l_keys_key_tl = \exp_not:n { #1 } }
 }
\clist_new:N \l_miguel_label_clist
\box_new:N \l_miguel_label_box
\box_new:N \l_miguel_label_image_box

\NewDocumentCommand{\xincludegraphics}{O{}m}
 {
  \group_begin:
  \tl_clear:N \l_miguel_label_tl
  \clist_clear:N \l_miguel_label_clist
  \keys_set:nn { miguel/label } { #1 }
  \tl_if_empty:NTF \l_miguel_label_tl
   {
    \miguel_includegraphics:Vn \l_miguel_label_clist { #2 }
   }
   {
    \SetHorizontalCoffin\imagecoffin
     {
      \miguel_includegraphics:Vn \l_miguel_label_clist { #2 }
     }
    \SetHorizontalCoffin\labelcoffin
     {
      \raisebox{\depth}
       {
        \bool_if:NTF \l_miguel_label_box_bool
         { \fcolorbox{white}{white}{\l_miguel_label_size_tl\l_miguel_label_tl} }
         { \l_miguel_label_size_tl\l_miguel_label_tl }
       }
     }
    \SetVerticalPole\imagecoffin{left}{3pt+\CoffinWidth\labelcoffin/2}
    \SetVerticalPole\imagecoffin{right}{\Width-3pt-\CoffinWidth\labelcoffin/2}
    \SetHorizontalPole\imagecoffin{up}{\Height-3pt-\CoffinHeight\labelcoffin/2}
    \SetHorizontalPole\imagecoffin{down}{3pt+\CoffinHeight\labelcoffin/2}
    \use:x{\JoinCoffins\imagecoffin[\l_miguel_label_pos_tl]\labelcoffin[vc,hc]} 
    \TypesetCoffin\imagecoffin
   }
   \group_end:
 }
\NewDocumentCommand{\setlabel}{m}
 {
  \keys_set:nn { miguel/label } { #1 }
 }

\cs_new_protected:Nn \miguel_includegraphics:nn
 {
  \includegraphics[#1]{#2}
 }
\cs_generate_variant:Nn \miguel_includegraphics:nn { V }

\ExplSyntaxOff

\begin{document}

\title{Do cosmological simulations reproduce the [OIII] 88 $\mu$m line emission and properties of JWST-discovered galaxies at $z \geq 12$?}

\correspondingauthor{Shengqi Yang}
\email{syang@carnegiescience.edu}

\author{Shengqi Yang}
\affiliation{Los Alamos National Laboratory, Los Alamos, NM 87545, USA}
\affiliation{Carnegie Observatories, 813 Santa Barbara Street, Pasadena, CA 91101, U.S.A}

\author{Adam Lidz}
\affiliation{Department of Physics and Astronomy, University of Pennsylvania, 209 South 33rd Street, Philadelphia, PA 19104, USA}

\author{Hui Li}
\affiliation{Los Alamos National Laboratory, Los Alamos, NM 87545, USA}

\author{Gerg\"o Popping}
\affiliation{European Southern Observatory (ESO), Karl-Schwarzschild-Strasse 2, Garching 85748, Germany}

\author{Jorge A. Zavala}
\affiliation{University of Massachusetts Amherst, 
710 North Pleasant Street, Amherst, MA 01003-9305, USA}

\author{Guochao Sun}
\affiliation{CIERA and Department of Physics and Astronomy, Northwestern University, 1800 Sherman Ave, Evanston, IL 60201, USA}

\begin{abstract}
Recent ALMA observations of the [OIII] 88 $\mu$m line provide spectroscopic confirmation of two
JWST photometric candidates, GS-z14 and GHZ2, at $z=14.2$ and $z=12.3$, respectively. These discoveries reveal that star formation and chemical enrichment were already underway when the universe was merely 300 Myr old, posing a challenge to galaxy formation models. Here we construct post-processed models for the [OIII] emission lines from galaxies in the state-of-the-art FIRE and IllustrisTNG simulations. Neither simulation suite contains galaxies directly comparable to GS-z14 or GHZ2. However, one simulated FIRE galaxy closely resembles GS-z14 in its star formation rate (SFR), stellar mass, metallicity, [OIII] luminosity and line-width, albeit at $z=8.7$, lagging GS-z14's formation by roughly 300 Myr. Although further investigation is required, we argue that the lack of simulated galaxies matching GS-z14 and GHZ2 may largely be a consequence of the limited volume of the FIRE simulations and the limited mass resolution of Illustris-TNG. We quantify the prospects for follow-up spectroscopic detections of GS-z14 in the [OIII] 52 $\mu$m line with ALMA, and in rest-frame optical [OIII] and Balmer lines with the MIRI instrument on JWST.  
\end{abstract}

\section{Introduction}

The \textit{James Webb Space Telescope (JWST)} has discovered a striking number of photometric
candidate galaxies at redshifts beyond $z \gtrsim 10$, pushing the galaxy formation frontier to within a few hundred million years after the Big Bang (e.g. \citealt{2022ARA&A..60..121R,2023ApJS..265....5H,2023ApJ...942L...9Y} and references therein). In fact, current estimates of the abundance of luminous ultraviolet (UV) photometric candidate galaxies at $z \gtrsim 10$ exceed the predictions of many models, while some JWST galaxies at slightly lower redshifts are estimated to have surprisingly large stellar masses (e.g. \cite{2023MNRAS.522.3986F,2024ApJ...969L...2F}).  
Importantly, recent ALMA observations of the [OIII] 88 $\mu$m line towards two JWST candidates, JADES-GS-z14-0 (hereafter GS-z14) and GHZ2, confirm high redshifts of $z=14.2$ and $z=12.3$, respectively, for these galaxies \citep{2024arXiv240920549S,2024NatAs.tmp..258Z,2024arXiv240920533C,2024ApJ...977L...9Z}.
In addition to providing spectroscopic confirmations, the [OIII] line detections allow metallicity estimates, probing early stages in the chemical enrichment history of our universe. 

Here we compare these striking observations with state-of-the-art cosmological simulations of galaxy formation, addressing the following questions. Do simulated galaxies reproduce the observed properties of GS-z14 and GHZ2, including their estimated star formation rates (SFRs), stellar masses, sizes, [OIII] luminosities and line-widths by $z \sim 12-14$? 
What are the formation histories for the closest simulated analogues to GS-z14 and GHZ2?
What are the prospects for follow-up emission line measurements towards GS-z14 and GHZ2? Can these better constrain the properties of the interstellar media (ISM) in these galaxies and reveal clues as to their formation mechanisms?

Specifically, we post-process ISM emission line models on top of state-of-the-art cosmological galaxy formation simulations and directly compare these with GS-z14 and GHZ2. 
Among the simulations employed here, the Feedback in Realistic Environments (FIRE) project provides tens of publicly available zoom-in simulations of galaxy formation during the Epoch of Reionization (EoR) at $z\gtrsim 5$. These simulations partly resolve the multi-phase ISM and include detailed sub-grid treatments for star formation and feedback \citep{2018MNRAS.478.1694M,2019MNRAS.487.1844M,2020MNRAS.493.4315M}. Encouragingly, previous work has found good agreement between the FIRE simulations and observational estimates of the mass-metallicity relationship (MZR) and UV luminosity functions during the reionization-era (e.g. \citealt{2023ApJ...955L..35S,2024ApJ...967L..41M, 2025MNRAS.536..988F}). 
Here, we use the open-source \textsc{HIILines} (\citealt{2023MNRAS.525.5989Y}, hereafter \citetalias{2023MNRAS.525.5989Y}) simulation framework to model emission lines on top of the simulated FIRE galaxies. 
In this framework, each FIRE-simulated star particle is assumed to be surrounded by a uniform and isolated HII region, whose properties are determined based on those of neighboring gas parcels. 
\textsc{HIILines} then calculates the equilibrium ionization structure, level populations, and resulting line luminosities (see also \citealt{2020MNRAS.499.3417Y}) in multiple oxygen and Balmer lines of interest from each HII region.  This modeling accounts for variations in the gas densities, metallicities, and stellar populations across each simulated galaxy. 

Further, we employ the approach of \cite{2024arXiv240903997Y} to make predictions for the line emission across the larger volume, yet lower resolution, IllustrisTNG simulations \citep{2019MNRAS.484.5587T,2020MNRAS.492.5167V}. Although the IllustrisTNG simulations do not resolve the multi-phase ISM, their larger volumes better capture the statistical properties of entire galaxy populations. Specifically, the work of \cite{2024arXiv240903997Y} uses a machine-learning technique, dubbed the ``Galaxy Mixture Density Network (GMDN)''. The GMDN determines the conditional probability distribution functions for the multi-line luminosity to stellar particle mass ratios around simulated star particles, given the stellar particle age, metallicity, and the galaxy's total stellar mass. The GMDN is trained on the FIRE zoom-in simulations and then applied to star particles from the IllustrisTNG simulation suite. Effectively, the FIRE simulations are used to calibrate 
a sub-grid line emission model which is applied on top of IllustrisTNG. 
In this work we assess whether the post-processed FIRE and IllustrisTNG models can reproduce the properties of GS-z14 and GHZ2. 

In the case of GHZ2, multiple rest-frame optical and UV lines have been successfully detected, providing important constraints on the ionizing spectrum and ISM properties for this galaxy \citep{2024ApJ...975..245C}, while only photometric observations have been obtained with MIRI for
GS-z14 \citep{2025NatAs.tmp...66H}. Here we quantify the prospects for follow-up spectroscopic observations of GS-z14 with MIRI and ALMA. We
discuss how bursty star formation leads to a large scatter in the [OIII] to UV luminosity ratio for the FIRE simulated galaxies and quantify
how this impacts the probability of [OIII] detections among UV-selected galaxies. 

The plan of this paper is as follows. In Section~\ref{sec:data} we describe the observed properties of GS-z14, GHZ2, and discuss the simulations used in this work. We present detailed comparisons between GS-z14, GHZ2, and simulated galaxies in Section~\ref{sec:compare}. In Section~\ref{sec:forecast} we discuss how bursty star formation can impact the detectability of [OIII] 88$\mu$m line emission from UV-selected galaxies. Finally, we present our conclusions in Section~\ref{sec:conclusion}.\par

\section{Data and Method}\label{sec:data}

GS-z14 is among the high-$z$ photometric candidates found by the JWST Advanced Deep Extragalactic Survey (JADES) Original Field program. This galaxy was discovered in the GOODS-S field \citep{2023arXiv231012340E,2024ApJ...970...31R}. Follow-up spectroscopy confirmed the presence of a Lyman-$\alpha$ break, and showed a tentative CIII]1907,1909\AA\ emission line doublet \citep{2024Natur.633..318C}, while subsequent ALMA observations detected the redshifted [OIII] 88 $\mu$m line at $6.6\sigma$ confidence \citep{2024arXiv240920549S,2024arXiv240920533C}. The ALMA observations pin-down the redshift of GS-z14 to $z=14.2$, making it the most distant galaxy that has been spectroscopically confirmed thus far. GS-z14 is also the third most UV-luminous galaxy among all of the 700 $z > 8$ JADES candidates. In addition to the galaxy's high UV luminosity, its [OIII] 88 $\mu$m emission requires rather metal-enriched gas (as discussed further below), suggesting that star formation started early and proceeded rapidly, at least in this extreme system \citep{2024A&A...689A.310F}. 

GHZ2 was discovered in the GLASS JWST Early Release Science Program \citep{2022ApJ...938L..15C,2022ApJ...940L..14N} as a photometric candidate galaxy at $z \sim 12$. Subsequent spectroscopy with NIRSpec/PRISM and MIRI/LRS confirmed its high redshift origin at $z=12.3$, placing it among the first pieces of evidence for a sizable abundance of UV-luminous galaxies at $z \gtrsim 10$ \citep{2024ApJ...972..143C}.
In the case of GHZ2, high-ionization transitions have been detected, including CIV and NIV] lines. This suggests that the emission from GHZ2 may partly arise from gas which is photo-ionized by an Active Galactic Nucleus (AGN), or by densely distributed low metallicity stellar populations \citep{2024ApJ...972..143C}. However, we will argue below that AGN contributions to the ionizing radiation in this source are likely subdominant. 
In addition to the high-ionization line measurements, there are detections of the Balmer-alpha line and emission lines from doubly-ionized oxygen in GHZ2. 
Specifically, \cite{2024NatAs.tmp..258Z,2024ApJ...977L...9Z} detected the [OIII] 88 $\mu$m line at $\sim5\sigma$, [OIII] 4960,5007\AA\ at $\sim9\sigma$, and H$\alpha$ at $\sim3.5\sigma$ significance from GHZ2. These measurements provide tight constraints on the gas-phase metallcity, density, and ionization parameter in this early galaxy \citep{2024ApJ...975..245C}. 
We summarize the detailed properties of GS-z14 and GHZ2 in Table~\ref{tb:GS-z14}. Note that GS-z14 is enriched to a metallicity of $Z \geq 0.14 Z_\odot$ already by $z=14.2$. 

\begin{table}[]
\centering
\begin{tabular}{|l|l|l|}
\hline
Parameter & GS-z14 & GHZ2         \\ \hline
redshift  & $14.1793(7)$  & $12.3327(35)$   \\ \hline
$M_\mathrm{UV}$   & $-20.81\pm 0.16$ & $-20.53\pm0.01$ \\ \hline
$\log M_*/[M_\odot]$    & $8.7^{+0.5}_{-0.4}$ & $9.05^{+0.10}_{-0.25}$               \\ \hline
SFR/[$M_\odot$/yr]       &     $25^{+6}_{-5}$ & $9\pm3$          \\ \hline
$L_\mathrm{[OIII],88}/[10^8L_\odot]$          &    $2.1\pm0.5$   & $1.3\pm0.3$         \\ \hline
FWHM/[km/s]  &   $136\pm31$   & $186\pm58$          \\ \hline
$R_{50}$/[pc]  &   $260\pm20$      & $105\pm9$       \\ \hline
$Z/[Z_\odot]$ & $>0.14$ &$0.05^{+0.12}_{-0.03}$ \\ \hline
\end{tabular}
\caption{
A summary of the properties of GS-z14 and GHZ2. The redshifts, absolute UV magnitudes (at a restframe wavelength of 1500 \AA), stellar masses, SFRs, [OIII] 88 $\mu$m line luminosities, the FWHMs of the [OIII] lines, the UV half-light radii, and the gas phase metallicities are given. This table is adapted from \cite{2024arXiv240920549S,2024NatAs.tmp..258Z,2024ApJ...977L...9Z}. The metallicities are determined using the method of \cite{2020MNRAS.499.3417Y}. The luminosities are corrected for gravitational lensing. 
}
\label{tb:GS-z14}
\end{table}

We compare GS-z14 and GHZ2 with the properties of the 22 FIRE high-$z$ suite galaxies (\citealt{2018MNRAS.478.1694M,2019MNRAS.487.1844M,2020MNRAS.493.4315M}, see \citealt{2018MNRAS.480..800H} for discussions of the simulation methodologies and underlying physics models in FIRE) and well-resolved IllustrisTNG galaxies \citep{2019MNRAS.484.5587T}, after post-processing with \textsc{HIILines} and the GMDN. Our earlier work showed that the 22 FIRE galaxies, post-processed with \textsc{HIILines} at $z=6$, are consistent with current ALMA and JWST line luminosity measurements, metallicities, SFRs, and other observable properties (\citetalias{2023MNRAS.525.5989Y}). Here we extend the 
\citetalias{2023MNRAS.525.5989Y}
analysis from $z=6$ out to higher redshifts, including earlier FIRE snapshots so that we can compare directly with GHZ2 and GS-z14 at $z \sim 12-14$. This extension also helps characterize the formation histories of the simulated galaxies. 

Note that throughout this work we assume that the main source of ionizing photons in both GS-z14 and GHZ2 is from stellar radiation, and neglect any AGN contributions. This assumption is supported by the extended restframe UV morphology of GS-z14 \protect\citep{2024arXiv240920549S}. In the case of GHZ2, a stellar radiation-dominated model is suggested by 
the low gas velocity dispersion, and the galaxy's extended morphology \protect\citep{2024Natur.633..318C}.
Further modeling efforts will be required to include any AGN contributions to the ionizing radiation field in these galaxies. \par 

\section{GS-z14 and GHZ2 versus simulations}\label{sec:compare}

\begin{figure*}
    \centering
    \includegraphics[width=0.99\textwidth]{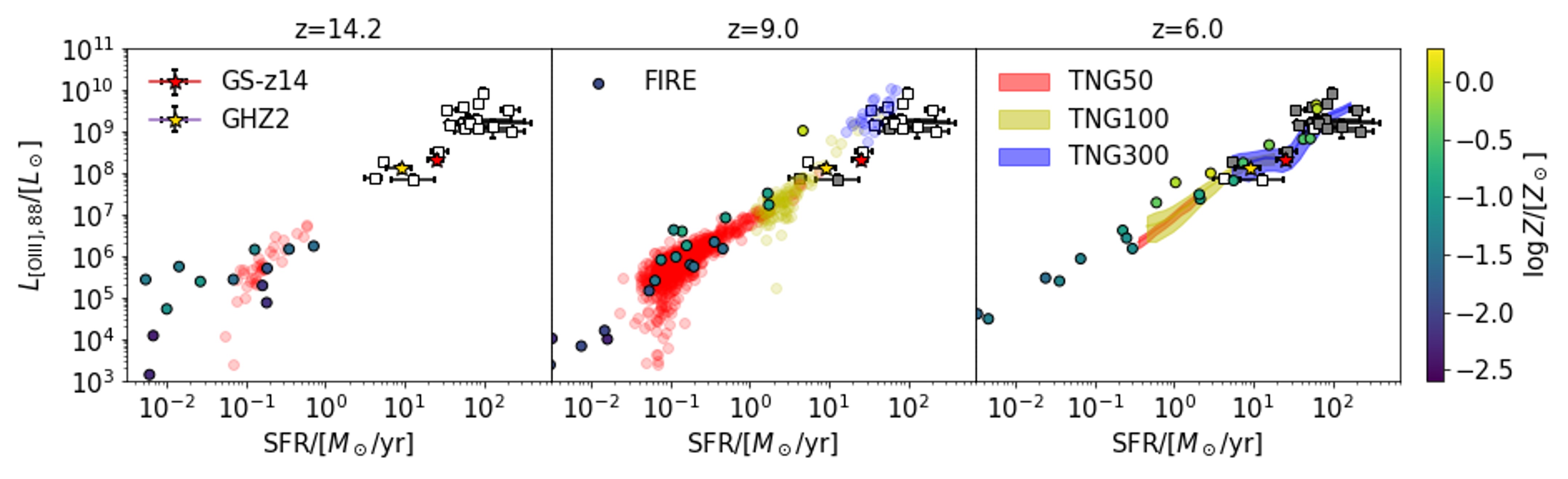}
    \caption{
A comparison between the observed [OIII] 88$\mu$m line luminosity versus SFR relation \citep{2018Natur.557..392H,2022MNRAS.515.1751W,2022ApJ...934...64A,2024MNRAS.527.6867A,2024ApJ...964..146F,2024arXiv240920549S,2024ApJ...977L...9Z}, FIRE simulations (\citetalias{2023MNRAS.525.5989Y}), and TNG galaxies \citep{2024arXiv240903997Y}. The shaded bands at $z=6$ give
the $1-\sigma$ range among TNG50, TNG100, and TNG300 simulated galaxies. At $z \geq 9$, the abundance of resolved TNG galaxies is small, and the results are shown as transparent points rather than bands. The observational measurements for GS-z14 and GHZ2 are shown as colored points with error bars, and are included in each panel to gauge redshift evolution. In each panel, observational results at nearby redshifts are included as gray points with error bars, while the white points show observations separated by more than $\Delta z = 1.5$ from the redshift of the simulations. The simulations reproduce the overall observed correlation between $L_\mathrm{[OIII],88}$ and SFR. However, neither simulation includes a galaxy with as high an SFR and [OIII] luminosity as GS-z14 or GHZ2 at $z \geq 12$. At later times, near $z=9$, the FIRE galaxy z5m12b has a line luminosity and SFR which are comparable to those of GS-z14 and GHZ2. }\label{fig:LOIIISFR}
\end{figure*}

Figure~\ref{fig:LOIIISFR} compares the $L_\mathrm{[OIII],88}$ versus SFR correlation in our simulated models with the observational results, including GS-z14, GHZ2, as well as those from lower redshift ALMA measurements in the EoR  \citep{2018Natur.557..392H,2022MNRAS.515.1751W,2022ApJ...934...64A,2024MNRAS.527.6867A,2024ApJ...964..146F,2024ApJ...977L...9Z}.
The simulated models include results for the 22 FIRE galaxies, post-processed with \textsc{HIILines}, and the IllustrisTNG galaxies after applying our sub-grid GMDN modeling. At $z=6$ the red, yellow, and blue bands show the $1-\sigma$ spread in the [OIII] luminosity-SFR correlation for TNG50, TNG100, and TNG300, respectively. At $z \geq 9$ the TNG galaxy abundance is smaller, and the results are shown as individual transparent colored points instead of bands. First, note that there is good general agreement between the $z \sim 6$ ALMA measurements and both the FIRE and TNG models at these redshifts, as discussed in \citetalias{2023MNRAS.525.5989Y} and \cite{2024arXiv240903997Y}. 

At higher redshifts, the simulations grow deficient in galaxies with large SFRs and [OIII] luminosities. At the observed redshift of GS-z14, $z=14.2$, the simulated galaxies with the largest SFRs have SFR $\sim 1 M_\odot/{\mathrm yr}$, more than an order of magnitude lower than that of GS-z14 and also well below that of GHZ2 at $z=12.3$. It is interesting to note, however, that the overall normalization of the $L_\mathrm{[OIII],88}$-SFR correlation does not appear to evolve strongly with redshift. In this sense, the main challenge of reproducing GS-z14 may be achieving the requisite SFR by $z \sim 14$, rather than producing the metal enrichment necessary to match the observed [OIII] emission. Of course this statement relies on extrapolating the simulated $L_\mathrm{[OIII],88}$-SFR relation over more than an order of magnitude in SFR. 

\par

\begin{figure}
    \centering
    \includegraphics[width=0.45\textwidth]{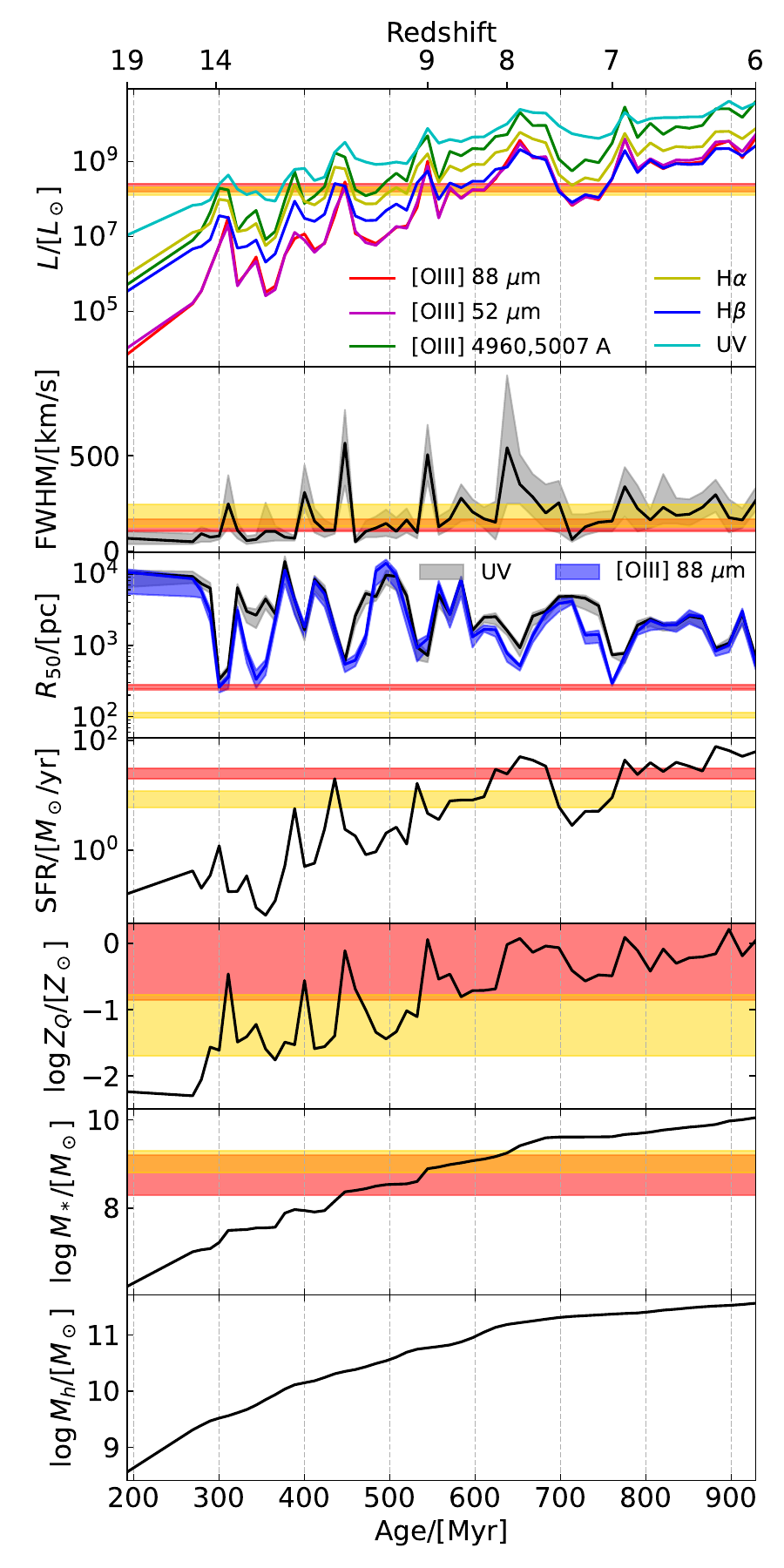}
    \caption{The formation and evolution of FIRE galaxy z5m12b.
    This simulated galaxy broadly resembles GS-z14, except it has similar properties at $z=8.7$ rather than at $z=14.2$. 
    The panels from top to bottom show (as a function of the age of the universe): the UV luminosity and those in various emission lines; the [OIII] 88 $\mu$m line-width (FWHM); the 2D UV and [OIII] half-light radii, $R_{50}$; the instantaneous SFR; the gas-phase metallicity weighted by the incident rate of hydrogen ionizing photons, $Z_Q$; the stellar mass, $M_*$; and the host dark matter halo mass, $M_{\rm h}$. In each panel, the red/gold horizontal bands indicate the $1-\sigma$ range in the observationally estimated properties for GS-z14/GHZ2  (Table~\ref{tb:GS-z14}). Note that in the top panel the colored bands give the measured [OIII] $88 \mu$m line luminosities.  
    The simulated FWHM and $R_{50}$ vary with viewing direction: the gray bands indicate their $1-\sigma$ spread. Note that the fluctuations in metallicity with time arise because the emission lines from HII regions probe photoionization rate-weighted metallicities (see text). Many of the properties of z5m12b near $z \sim 9$ match those of GS-z14, although $R_{50}$ is usually larger in simulations than observations.}\label{fig:z5m12b_t}
\end{figure}

Although none of the simulated galaxies directly reproduce the properties of GS-z14, one FIRE galaxy, z5m12b, matches it in [OIII] luminosity and SFR near $z \sim 9$. Since this simulated galaxy is a potential later-forming analogue to GS-z14, lagging the observed galaxy by roughly 300 Myr, it is interesting to examine its growth history. This is shown in Figure~\ref{fig:z5m12b_t}. 

Specifically, this figure shows the time evolution of (from top to bottom): the simulated UV luminosity and various line luminosities, the [OIII] 88 $\mu$m line full-width-half-maximum (FWHM), UV and [OIII] 88 $\mu$m emission 2D half-light radii ($R_{50}$), the SFR, the gas-phase metallicity weighted by the rate of hydrogen ionizing photon production ($Z_{\mathrm Q}$), the stellar mass, and the host dark matter halo mass of z5m12b. The UV continuum luminosities are computed using the spectra from an FSPS population synthesis model \citep{2009ApJ...699..486C}, summed over all stellar particles, and integrated over restframe wavelengths from 1450\AA\ to 1550\AA\ using a top-hat filter. We define the FIRE galaxy center as the [OIII] 88 $\mu$m luminosity-weighted stellar particle center. The 2D $R_{50}$ is then defined as the radius within which the enclosed stellar particles (i.e. with $x,y$ positions such that $x^2+y^2\leq R_{50}^2$) contribute half of the galaxy-wide UV/[OIII] 88 $\mu$m luminosity. We assume that the observational estimates of the half-light radii and line-widths fully account for uncertainties in the beam deconvolution and spectral resolution. 
The [OIII] line-widths and the [OIII]/UV half-light radius are sensitive to the viewing direction, and so we randomly select 100 orientations for the line-of-sight (LOS) direction. 
The resulting median line FWHM and $R_{50}$ results are shown with blue and black curves, while the gray and blue bands give the $1-\sigma$ spread across different viewing directions. The red/gold bands show the estimated properties for the observed galaxies, GS-z14/GHZ2, with the widths of these bands indicating $1-\sigma$ confidence intervals.  

As illustrated in Figure~\ref{fig:z5m12b_t}, z5m12b has a bursty star formation history in which the SFR can jump by up to an order of magnitude within 10 Myrs. The HII region line luminosities quickly rise following a burst of star formation, as does the UV luminosity, although this is
sensitive to the star formation rate averaged over a longer time scale. The galaxy can hence shift above and below the detectability thresholds in luminosity
over time \citep{2023MNRAS.526.2665S,2023ApJ...955L..35S}. This FIRE galaxy first achieves a UV luminosity, SFR, [OIII] luminosity, and stellar mass that are comparable to GS-z14 near $z \sim 10$, following a massive starburst event, approximately 450 Myr after the Big Bang. Subsequently, z5m12b undergoes further growth and star formation, eventually exceeding the [OIII] luminosities of both GS-z14 and GHZ2 below $z \lesssim 8$ or so.
Following each simulated starburst, the resulting young stellar populations doubly-ionize oxygen, sourcing [OIII] emission. The gas-phase metallicity gradually builds up as supernova explosions enrich the surrounding gas. However, the ionizing photon-weighted metallicity probed by the [OIII] lines fluctuates in time, with declines as the massive and short-lived, ionizing photon-producing, stars die following a starburst. 

Figure~\ref{fig:z5m12b_t} also compares the [OIII], H$\alpha$, and H$\beta$ line luminosities. This reveals how much of the luminosity evolution is driven by increases in the intensity of ionizing radiation as opposed to evolution in the
metallicity of the HII regions. Specifically, the [OIII] line luminosities are sensitive to variations in both the metallicity and the rate of ionizing photon production, while the Balmer lines depend only on the latter quantity. At $z \gtrsim 8$, the simulations show faster evolution in the [OIII] luminosities than in the Balmer lines, while at lower redshift the [OIII] and Balmer lines increase at similar rates. This arises because at high redshift the [OIII] luminosity evolution is driven primarilly by metallicity growth, while the similar trends across different lines at
lower redshift indicate that the [OIII] evolution mainly traces evolution in the strength of the ionizing radiation field.

The [OIII] FWHM can vary by as much as a few hundred km/s when the galaxy is viewed from different directions. This arises in part because the simulated galaxy has undergone recent mergers, and so it contains multiple components with sizable relative velocities, and each component may host HII regions with [OIII] emitting gas. The FWHM may also partly reflect outflowing and inflowing gas. The simulated FWHMs are broadly consistent with the observed ones,  although the simulations tend to exceed the measured values close to starburst events. 

The half-light radii, $R_{50}$, are determined observationally from the restframe UV light, and
it is assumed that $R_{50}^\mathrm{[OIII]}=R_{50}^\mathrm{UV}$ for dynamical mass determinations. The third panel of Figure~\ref{fig:z5m12b_t} demonstrates that the two radii generally match. The short-lived massive stars, which produce the ionizing photons that source the [OIII] emission, tend to be produced in concentrated starburst events, while the UV light receives some contributions from lower-mass stars which generally have a more extended spatial distribution. Consequently, $R_{50}^\mathrm{[OIII]}$ is usually a little smaller than $R_{50}^\mathrm{UV}$. The figure also reveals that $R_{50}$ tends to grow after a starburst event. This may be a consequence of gas shocked by supernova feedback: as a supernova blast wave sweeps through the surrounding ISM it compresses gas and triggers extended star formation. Finally, in some cases, following merger events the size estimates reflect the emission from multiple different components. The typical $R_{50}$ values in the simulations exceed the observational results in GS-z14 by a factor of $1.3-60$, where the broad range here reflects the spread across viewing directions. 
Although the [OIII] line-widths can be used to determine a reasonably accurate dynamical mass for z5m12b,
the large $R_{50}$ in our models leads to a simulated dynamical mass which is $\sim 10$ times greater than the observed value of
$M_\mathrm{dy}=10^9M_\odot$ value estimated by \cite{2024arXiv240920549S}.

The difference in $R_{50}$ between the simulations and observations is still more pronounced for GHZ2, whose observed size is only $R_{50} \sim 100$ pc. We do not find {\em any} viewing directions towards z5m12b where the half-light radius is so small. 
The reason for this discrepancy is unclear. It might owe to our comparison between GS-z14/GHZ2 and a lower redshift simulated analogue: a proper $z \sim 12-14$ simulated galaxy with the same SFR may be more compact. On the other hand, the simulated galaxies do roughly match the observed $R_{50}$ value for GS-z14 along rare viewing angles. Alternatively, this could indicate an observational bias: for example, the observations may preferentially select only the brightest star-forming clumps and miss more extended emission components. That is, the observational size determinations may miss the low surface brightness outskirts of the galaxies, although these are included in our theoretical calculations. Likewise, dust obscuration effects could complicate the observational size estimates. 
Finally, the observed sizes could suffer from an inaccurate determination or deconvolution of the point-spread function \citep{2022A&A...659A.141S}. 

\par

\begin{figure}
    \centering
    \includegraphics[width=0.45\textwidth]{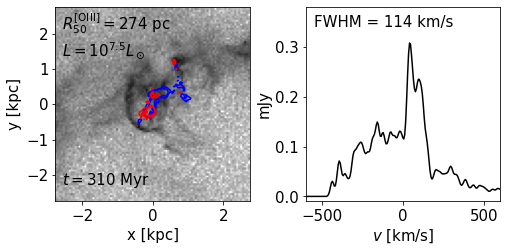}\\
    \includegraphics[width=0.45\textwidth]{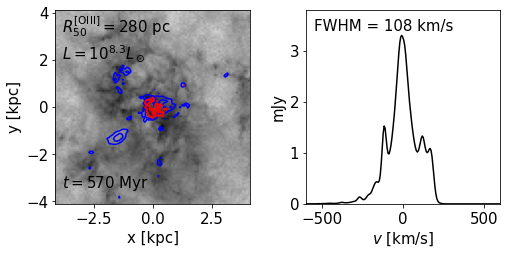}
    \caption{Visualizations of z5m12b at $z=13.5$ (top) and $z=8.7$ (bottom). The LOS direction is selected such that the [OIII] emission half-light radius and FWHM are consistent with GS-z14. In the left panel of each figure the [OIII] surface brightness, the UV continuum surface brightness at 1500\AA, and the gas distribution are shown in the red contours, blue contours, and the gray map, respectively. The right sub-panels show the [OIII] 88 $\mu$m line profile.}\label{fig:z5m12b_visual}
\end{figure}

Figure~\ref{fig:z5m12b_visual} presents visualizations of the gas distribution, UV light, and [OIII] emission from the simulated FIRE galaxy z5m12b. For comparison, the top panels show results at $z=13.5$, while the bottom panels are at $z=8.7$. The $z=13.5$ case is chosen since this snapshot follows a recent starburst, and is close to the redshift of GS-z14, although the simulated galaxy at this redshift underproduces GS-z14's SFR and [OIII] luminosity, the latter by about 0.8 dex. On the other hand, z5m12b matches the [OIII] luminosity at the redshift of the bottom panel, $z=8.7$. In both the top and bottom panels, the viewing direction has been chosen so that the simulated galaxies match the observed values of $R_{50}$ and the FWHM of the [OIII] 88 $\mu$m line. As discussed earlier, reproducing $R_{50}$ requires an atypical viewing direction. The red and blue contours show the $1-\sigma$ and $2-\sigma$ range in simulated surface brightness, while the gray map shows the projected gas mass distribution from the simulations. In each case, the [OIII] and UV surface brightness distributions are similar and more compact than the overall gas distribution, since the emission traces only dense regions in the ISM. 
Finally, the right panels show the simulated [OIII] 88 $\mu$m emission line profiles. Along the viewing directions chosen, the simulated FWHMs match the observational estimates for GS-z14 and GHZ2. (Note that the normalization of the line profile in the $z=8.7$ snapshot is larger than observed because of the simulated galaxy's lower redshift).\par 

A recent ALMA follow-up study places an upper bound on the [CII] 158 $\mu$m signal from GS-z14 \citep{2025arXiv250201610S}. That work finds that the [OIII]/[CII] ratio must be larger than in local galaxies, in line with previous results among low SFR galaxies at $6 < z < 9$. The authors further find that the ISM gas density within GS-z14 should lie between $20 < n_\mathrm{H}/[\mathrm{cm}^{-3}]<125$. We
caution, however, that the `one-zone' model adopted to determine the gas density may lead to misleading conclusions \citep{2023MNRAS.525.5989Y}. In our post-processed FIRE calculations, for example, the gas density variations across each galaxy are sizable and this complicates the interpretation of
line luminosity measurements. In the case of [OIII] emission, the [OIII] luminosity-weighted gas density in z5m12b at $z=8.7$ is 250 cm$^{-3}$, which lies above the observational bound, even though the average gas density across all HII regions in this simulated galaxy is 64 cm$^{-3}$ in coincidental agreement with the results of \citet{2025arXiv250201610S}. Moreover, the typical [CII] luminosity-weighted density probed by [CII] observations may differ from that traced by [OIII] lines. Indeed, it is likely that high redshift [CII] emission mainly comes from photo-dissociation regions with $T \sim 100$ K \citep{2019MNRAS.489....1F,2024A&A...682A..98S}. If these regions are in approximate pressure equilibrium with the warmer HII regions, then they are necessarily at higher density. At any rate, further modeling efforts will be required to include [CII] in our modeling framework to help interpret the interesting new [CII] luminosity bounds from GS-z14. \par
Another recent follow-up analysis constructs velocity maps from the GS-z14 [OIII] 88 $\mu$m ALMA measurements \citep{2025arXiv250310751S}. This study tentatively finds a velocity gradient, and argues that the galaxy contains a cold, rotationally-supported disk. This contrasts with the galaxies in the FIRE simulations which do not show well-formed disks at $z \geq 6$. Further higher-resolution observations and comparisons with full mock ALMA data constructed from our models will be required to understand
whether there is a genuine discrepancy here.
 
\par

The simultaneous detection of [OIII]4960,5007\AA, [OIII] 88 $\mu$m, and [OII] 3726,29\AA\ lines from GHZ2 allows tight constraints on the average HII region gas-phase metallicity, gas temperature, and on the formation history of this galaxy \citep{2024ApJ...975..245C, 2024MNRAS.534.3665Y}. Given the consistency between the properties of z5m12b at $z=8.7$ and GS-z14, we can use it to predict how luminous GS-z14 should be in additional emission lines and forecast their detectability with ALMA or MIRI. Specifically, it may be interesting to search for [OIII] 52 $\mu$m, [OIII] $4960,5007$\AA, H$\alpha$ and H$\beta$ line emission from this galaxy. 
This would potentially help sharpen recent photometric observations of GS-z14 with MIRI \citep{2025NatAs.tmp...66H}. 
The predicted luminosities in these lines for GS-z14 are summarized in Table~\ref{tb:prediction}. The [OIII] 4960,5007\AA/H$\beta$ ratio in z5m12b is about 7, and is consistent with the typical [OIII]/H$\beta$ ratios among the 
the JADES Data Release samples at $z\approx8$ \citep{2025ApJS..277....4D}. 
However, \cite{2025NatAs.tmp...66H} uses their photometric detections of GS-z14 and spectral energy distribution (SED) models to find
a lower [OIII]/H$\beta$ ratio (see their Extended Data Figure 5 for details), which they argue indicates a metallicity of  $Z\sim0.02Z_\odot$. This metallicity is hard to reconcile with the [OIII] 88$\mu$m luminosity from GS-z14, which we find implies $Z \geq 0.14 Z_\odot$ (Table~\ref{tb:GS-z14}). Spectroscopic observations of GS-z14 with MIRI may provide a cross-check here. 
\par
Unfortunately, at $z=14.2$ the [OIII] 52 $\mu$m line is redshifted to a frequency which lies in between ALMA's band 7 and band 8 and where there is negligible transmission through the Earth's atmosphere. However, the predicted [OIII] 4960,5007\AA\ line flux is $2.9\times10^{-18}$ $\mathrm{erg/s/cm^2}$ at $z=14.2$, only a factor of $\sim 2$ smaller than that detected in the case of GHZ2 at $\sim9\sigma$ significance. Therefore, encouragingly, a similar 9-hour observing strategy with MIRI should allow detections of these [OIII] lines at a signal-to-noise-ratio (SNR) of $\sim 5-6$ for GS-z14. In the case of H-$\alpha$, the JWST exposure time calculator suggests that an 18-19 hour observation is required to detect this line at an SNR of 5 with MIRI. Unfortunately, several dozen hours will be required to detect H-$\beta$ from this source. As mentioned in Section~\ref{sec:data}, the [OIII]4960,5007\AA, [OIII] 88 $\mu$m, [OII] 3727,29\AA, and H$\alpha$ lines have been detected with high significance by JWST/MIRI and ALMA for GHZ2 \citep{2024ApJ...975..245C}. We use a new strong line diagnostic method which combines semi-analytical [OIII], [OII] line emission models with empirical gas temperature-metallicity correlations (see \citealt{2024MNRAS.534.3665Y} for details regarding our strong line diagnostic analysis methods) to derive the metallicity of this galaxy. For GHZ2, the measurements of $(L^\mathrm{[OIII]}_{4960,5007}+L^\mathrm{[OII]}_{3727,29})/\mathrm{H}\beta$ and $L^\mathrm{[OIII]}_{4960,5007}/L^\mathrm{[OII]}_{3727,29}$ imply a metallicity of $Z = 0.06_{-0.03}^{+0.11}Z_\odot$. This result is consistent with the more empirical strong line diagnostic approach of \citep{2024ApJ...962...24S,2024ApJ...975..245C}, and is generally lower than the metallicity of z5m12b at $z\leq9$.

\begin{table}[]
\centering
\begin{tabular}{|l|c|c|c|}
\hline
Line & L [$10^8 L_\odot$] & F [Jy km/s] & $t$ [hour]\\ \hline
[OIII] 52 $\mu$m & $2.4\pm 0.6$ & $(1.6\pm0.4)$E-3 & $^\dagger$        \\ \hline
[OIII] 4960,5007\AA & $22\pm5$ & $(1.4\pm0.3)$E-4 &  $\sim10$      \\ \hline
H$\alpha$ & $9\pm2$ & $(7.6\pm1.7)$E-5 & $\sim20$       \\ \hline
H$\beta$ & $3.1\pm0.7$ & ($2.0\pm0.4$)E-5& $>$100         \\ \hline
\end{tabular}
\caption{Predictions for GS-z14's luminosities, fluxes, and the observing times required for $5\sigma$ detections in additional emission lines. The predictions are based on model line ratios from the FIRE galaxy z5m12b at $z=8.7$, which matches many of the observed properties of GS-z14. $^\dagger$At this redshift, this transition is not observable from the ground.}
\label{tb:prediction}
\end{table}

\section{Detection rate estimated from TNG}\label{sec:forecast}

As mentioned earlier, the star formation history in Figure~\ref{fig:z5m12b_t} varies on rather short time scales: z5m12b and other FIRE galaxies show ``bursty'' star formation \citep{2023MNRAS.526.2665S,2023ApJ...955L..35S}. The luminosity versus time behavior in the top panel of this figure further shows that the [OIII] line luminosity is sensitive to short $\sim 10$ Myr timescale variations in the star formation rate, while the smoother UV luminosity curve reflects the SFR averaged over longer timescales. This behavior arises because the HII regions where the [OIII] emission originates are sourced by ionizing photons from short-lived massive stars, while the UV continuum emission receives contributions from older stellar populations. This difference leads to a sizable scatter in the [OIII] luminosity for galaxies with a given absolute magnitude in the UV. Here we quantify the impact of this scatter on efforts to conduct follow-up searches for [OIII] 88$\mu$m emission around UV-selected galaxies. This is a potentially important contributing factor in explaining {\em non-detections} of the [OIII] 88$\mu$m line around JWST-selected EoR candidate galaxies. 

\begin{figure}
    \centering
    \includegraphics[width=0.45\textwidth]{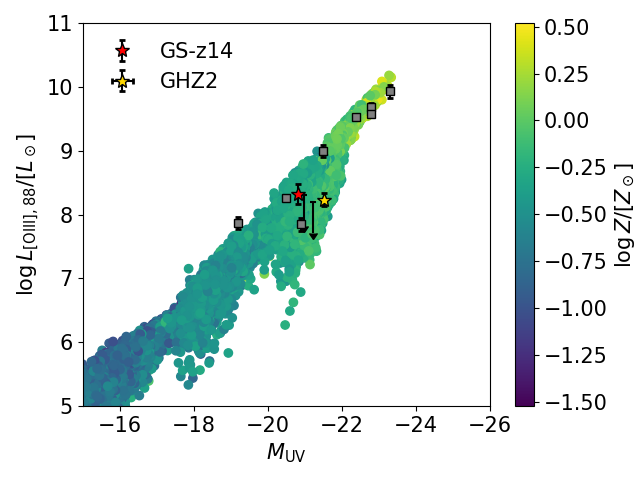}
    \caption{
    A comparison between simulated and observed [OIII] 88 $\mu$m luminosities, as a function of absolute UV magnitude. The colored points show TNG galaxies, color-coded by their metallicities, modeled with our GMDN-based approach. The gray data points show current ALMA results, with the downward pointing arrows indicating non-detections of the 88 $\mu$m line \citep{2021A&A...646A..26B,2023ApJ...950...61Y}. The large scatter at fixed UV magnitude among the simulated galaxies is beyond that expected from metallicity variations alone and partly owes to bursty star formation.}\label{fig:TNG}
\end{figure}

Figure~\ref{fig:TNG} illustrates the issue using our GMDN-processed IllustrisTNG galaxies at $z=6$, in comparison with current measurements.\footnote{Note that due to the relatively low resolution, IllustrisTNG simulations can not resolve the clustering of supernovae in the multi-phase ISM, and do not show the same bursty star formation observed in the FIRE simulations. However, our sub-grid line emission model is derived from FIRE and so when we add line emission to the TNG galaxies, the model effectively inherits the stochastic luminosity variations in FIRE.}
The UV luminosities are attenuated according to dust model A from 
\cite{2020MNRAS.492.5167V}. The simulated results are in good agreement with current observations. Note that while there are substantial fluctuations in the average gas-phase metallicity from galaxy to galaxy, this does not fully account for the large 
$L_\mathrm{[OIII]}$ scatter at fixed $M_{\rm UV}$. For example, the $\log L_\mathrm{[OIII]}$ and $\log Z$ distributions for galaxies with $-20.5\leq M_\mathrm{UV}\leq-20$ are close to Gaussian. The standard deviation of $\log L_\mathrm{[OIII]}$ is $\sim0.3$ dex. However, the metallicity scatter is only $\sim0.1$ dex. This partly owes to the bursty star formation effects, which can then explain some of the non-detections of the $88 \mu$m line in the literature. 

Quantitatively, we can assess the detectability of each simulated TNG galaxy in the [OIII] 88$\mu$m line as a function of its UV luminosity. For simplicity, we fix the simulated luminosities to their $z = 6$ values and consider their detectability at various different redshifts. In each case, we set a $5-\sigma$ detection threshold assuming 10 hours of integration with ALMA and a 200 km/s FWHM for the 88 $\mu$m line. 
The fraction of detectable galaxies is well described by a reverse sigmoid function:
 \begin{equation}\label{eq:f_detect}
     f_\mathrm{detection}=1-\dfrac{1}{1+\exp(-aM_\mathrm{UV}-b)}\,.
 \end{equation}
Given the scatter, the chance of detecting the [OIII] 88 $\mu$m line from a galaxy reaches 50\% when the galaxy's UV magnitude is
$M_{\rm UV} = -b/a$. The detection rate varies smoothly between $10\%$ and $90\%$ for $M_{\rm UV} = -b/a + (2.2/a)$ to
$M_{\rm UV} = -b/a - (2.2/a)$. Beyond this range, the detection probability falls rapidly to zero at the faint end, while it quickly reaches unity for more UV-luminous galaxies. However, the sizable [OIII] luminosity scatter at fixed $M_{\rm UV}$ means that only rather UV bright galaxies make [OIII] detections highly probable. In other cases, a relatively UV bright galaxy may result from a fading starburst yielding only weak [OIII] emission.   

\begin{table}[]
\begin{tabular}{|l|l|l|l|}
\hline
redshift & flux {[}Jy km/s{]} & a    & b    \\ \hline
6.10     & 0.157              & 3.54 & 75.6 \\ \hline
7.10     & 0.104              & 3.21 & 68.0 \\ \hline
8.15     & 0.0878             & 3.22 & 68.4 \\ \hline
9.00     & 0.0399             & 4.52 & 93.2 \\ \hline
10.00    & 0.0397             & 4.34 & 90.0 \\ \hline
11.00    & 0.0412             & 3.70 & 77.2 \\ \hline
12.00    & 0.0311             & 4.33 & 89.7 \\ \hline
13.00    & 0.0369             & 3.36 & 70.4 \\ \hline
14.00    & 0.0304             & 3.64 & 75.9 \\ \hline
\end{tabular}
\caption{
Parameter fits for Equation~\ref{eq:f_detect}, characterizing
the probability that a galaxy with a given absolute UV magnitude is detectable in the [OIII] 88 $\mu$m line. The second column
gives the minimum integrated line flux for a $5-\sigma$  detection with ALMA after 10-hours of integration time. These results assume an [OIII] line FWHM of 200 km/s. }
\label{tb:f_detect}
\end{table}

 \par

\section{Discussion and Conclusions}
\label{sec:conclusion}

The high redshifts, large star formation rates, stellar masses, compact sizes, and high [OIII] luminosities of GS-z14 and GHZ2 present a challenge to models of early galaxy formation. Here, after modeling the line emission around 22 FIRE zoom-in galaxies and simulated galaxies from the IllustrisTNG simulations, we find that neither simulation successfully reproduces the properties of either GS-z14 or GHZ2. 

In the case of the FIRE simulations, one possibility is that they simply lack the volume required to capture rare sources. Note that the FIRE zoom-in simulations are drawn from regions in a set of cosmological boxes which are each smaller than 43 cMpc on a side. On the other hand, the co-moving volume spanned by the JADES survey (used to find GS-z14) from $z \sim 10-15$ is $\sim(100\ \mathrm{ cMpc})^3$, more than ten times bigger than the largest FIRE box from which the publicly available zoom-in regions are drawn. In order to investigate the impact of the limited FIRE volumes, we first extrapolate the $L_{\rm [OIII]}-{\rm SFR}$ correlation at $z=14$ in FIRE out to larger SFRs than directly captured in the simulations. We then combine this with the SFR-UV correlation in \cite{2016MNRAS.460..417S} and the UV luminosity function estimated from the FIRE-2 simulations in 
\cite{2023ApJ...955L..35S} at the nearest available redshift of $z \sim 12$. Finally, we assume that the $L_{\rm [OIII]}-{\rm SFR}$ correlation follows a lognormal distribution with a scatter of 0.2 dex, as suggested by the TNG simulation results at $z=6$ \citep{2024arXiv240903997Y}. 
The resulting [OIII] 88 $\mu$m luminosity function may be fit by a Schechter function with:
 \begin{equation}
     \phi(L)=\phi^*\left(\dfrac{L}{L_*}\right)^\alpha\exp\left(-\dfrac{L}{L_*}\right)\dfrac{1}{L_*}\,.
 \end{equation}
 Here $L$ is the [OIII] 88 $\mu$m line luminosity in units of $L_\odot$, while $\phi_*=10^{-3.77}$ cMpc$^{-3}$, $L_*=10^{7.61}L_\odot$, and $\alpha=-1.70$. 
 This is only a rough estimate as it requires extrapolating the simulated $L_{\rm [OIII]}-{\rm SFR}$, and it ignores redshift evolution in the UV luminosity function fit from $z \sim 12$ to $z \sim 14$. It also considers larger UV luminosities than directly captured in the simulations of \cite{2023ApJ...955L..35S}. Nevertheless, based on this estimate it is interesting to note that we expect one galaxy with luminosity equal to or larger than that of GS-z14 in a volume of $(280\, {\rm cMpc})^3$ for GS-z14's best fit [OIII] 88 $\mu$m line luminosity, or $(160\, {\rm cMpc})^3$ if we adopt the $1-\sigma$ lower bound on the luminosity (see Table~\ref{tb:GS-z14}) at z=14. As this is comparable to the relevant JADES survey volume, it appears plausible that the difficulty of reproducing GS-z14 and GHZ2 reflects the challenge of capturing rare sources in simulations rather than indicating a defect in the treatment of galaxy formation physics in FIRE. In this regard, it is encouraging that the FIRE galaxy z5m12b matches many of GS-z14's properties at later cosmic times (by $\sim 300$ Myr) and it may be that galaxy formation is more advanced in large-scale overdense regions which are hard to sample in small simulation volumes. 
 Of course, this explanation is a bit unsatisfying, and further explorations will be required to draw more definitive conclusions. 
 
 Alternatively, the publicly available FIRE high-$z$ suite may underestimate the earliest phases of star and galaxy formation. For instance, the simulations may fail to capture feedback-free starburst modes which might lead to higher star formation efficiencies among $z \gtrsim 10$ galaxies \citep{Dekel23}.  It is also conceivable that the mismatch points to an inadequate modeling of still earlier phases in the star formation history of the simulated $z \sim 14$ galaxies: this may require  
 careful Pop-III star formation models and a detailed treatment of metal-free molecular hydrogen line cooling.  

 Regarding the simulation volume issue, one might naively suppose that the larger volume TNG simulation suite would better capture rare galaxies like GS-z14 and GHZ2, at least when supplemented with our sub-grid line emission model. However, these galaxies also have relatively small stellar masses which can be challenging to resolve in large volume simulations. These small galaxies are nevertheless rare because they form at high redshifts, $z \sim 12-14$, where even relatively small mass peaks in the density distribution lie on the exponential tail of the halo mass function, at least in currently favored cold dark matter cosmological models. Hence, capturing analogues of these galaxies directly in simulations requires both large volume and high mass resolution. 
 For example, TNG300 fails to resolve galaxies with stellar mass smaller than $M_* \lesssim 10^9 M_\odot$ galaxies, yet this would be necessary to capture galaxies like GS-z14 and GHZ2. 
 Zoom-in simulations on rare overdense peaks in the density distribution, where galaxy formation presumably occurs first, may be instructive. 

Since the FIRE galaxy z5m12b at $z=8.7$ matches GS-z14's main properties, we use it to predict GS-z14's [OIII] 52 $\mu$m, [OIII] 4960,5007\AA, H$\alpha$ and H$\beta$ line luminosities to
assist ALMA and JWST/MIRI follow-up measurement efforts. Among these, we find that JWST/MIRI can detect the [OIII] 4960,5007\AA\ lines with an integration time of less than 10 hours, while $\sim 20$ hours of observing time are required to detect H$\alpha$ from this source. Although these estimates indicate that further line detections from this current redshift record-holder are challenging, such measurements should be valuable. Specifically, the additional [OIII] line detections will help determine the typical gas temperatures in this galaxy's HII regions, while an H$\alpha$ measurement would help further pin-down the gas-phase metallicity when combined with [OIII] line measurements. The H$\alpha$ luminosity would also yield information regarding the rate of production of ionizing photons in GS-z14. 
These measurements, and related future discoveries, combined with the additional simulation efforts suggested above, will help more definitively assess the level of agreement between current galaxy formation simulations and observations of the first galaxies. 

\section{Acknowledgements}
We thank Christopher C. Hayward for sharing the full catalog of z5m12b at $z\geq6$ with us. S.Y. acknowledges support from the Director’s postdoctoral fellowship funded by the Laboratory Directed Research and Development (LDRD) program of Los Alamos National Laboratory (LANL) under project No. 20240863PRD2. H.L. acknowledges the support by LANL's LDRD project number 20220087DR. G.S. was supported by a CIERA Postdoctoral Fellowship.

\bibliography{z14}{} 
\bibliographystyle{aasjournal}

\end{document}